\documentclass[journal]{IEEEtran}
\usepackage{amssymb}
\ifCLASSINFOpdf
  \usepackage[pdftex]{graphicx}
\else
  \usepackage[dvips]{graphicx}
\fi

\usepackage[cmex10]{amsmath}
\usepackage{algorithm}
\usepackage{algorithmic}

\usepackage{array}

\usepackage{mdwmath}
\usepackage{mdwtab}

\usepackage{caption}
\usepackage{subcaption}
\usepackage{subfig} 
\usepackage{url}

\begin{document}

\title{Fair Scheduling for Data Collection in Mobile Sensor Networks with Energy Harvesting}

\author{Kai~Li,~\IEEEmembership{Member,~IEEE,}
        Chau~Yuen,~\IEEEmembership{Senior Member,~IEEE,}
        Branislav~Kusy,~\IEEEmembership{Senior Member,~IEEE,}
        Raja~Jurdak,~\IEEEmembership{Senior Member,~IEEE,}
        Aleksandar~Ignjatovic,         
        Salil~S.~Kanhere,~\IEEEmembership{Senior Member,~IEEE,}
        and Sanjay~Jha,~\IEEEmembership{Senior Member,~IEEE,}        
\thanks{An earlier version of this article appeared in the Proceedings of the 11th the European Conference on Wireless Sensor Networks (\textit{EWSN}) [Li et al. 2014]. This article features a new scheduling optimisation with energy harvesting, experimental characterisation and more complete theoretical analysis. 

K.~Li and C.~Yuen are with The Singapore University of Technology and Design (SUTD), Singapore. (e-mail: kai\_li@sutd.edu.sg) 

B.~Kusy and R.~Jurdak are with Autonomous Systems Lab, ICT Centre, CSIRO, Australia. 

A.~Ignjatovic, S.~S.~Kanhere, and S.~Jha are with the School of Computer Science and Engineering, The University of New South Wales, Australia.}% <-this % stops a space
}

\IEEEcompsoctitleabstractindextext{%
\begin{abstract}
\boldmath
We consider the problem of data collection from a continental-scale network of energy harvesting sensors, applied to tracking mobile assets in rural environments. Our application constraints favour a highly asymmetric solution, with heavily duty-cycled sensor nodes communicating with powered base stations. We study a novel scheduling optimisation problem for energy harvesting mobile sensor network, that maximises the amount of collected data under the constraints of radio link quality and energy harvesting efficiency, while ensuring a fair data reception. We show that the problem is NP-complete and propose a heuristic algorithm to approximate the optimal scheduling solution in polynomial time. Moreover, our algorithm is flexible in handling progressive energy harvesting events, such as with solar panels, or opportunistic and bursty events, such as with Wireless Power Transfer. We use empirical link quality data, solar energy, and WPT efficiency to evaluate the proposed algorithm in extensive simulations and compare its performance to state-of-the-art. We show that our algorithm achieves high data reception rates, under different fairness and node lifetime constraints.%, which limits available energy for wireless data transfer. 
\end{abstract}

}

\maketitle

\IEEEdisplaynotcompsoctitleabstractindextext
\IEEEpeerreviewmaketitle

\section{Introduction}
\label{intro}
\IEEEPARstart{R}{ecent} advances in embedded systems and battery technology have enabled a new class of large-scale wireless sensing applications~\cite{li2014kappa,liu2011self}. Consider a swarm of micro-aerial vehicles fitted with a variety of sensors that can achieve fine-grained three-dimensional sampling of our physical spaces, enabling exciting new applications such as urban surveillance, disaster recovery and environmental monitoring~\cite{Dantu2011,sinha2009multi}. It is now possible to monitor individual movement patterns of wildlife alongside the various aspects of their environment~\cite{dyo2012wildsensing,israel2011uav,jurdak2010adaptive,Dyo2010}. In a typical mobile sensing scenario, sensor nodes mounted on a carrier (e.g., vehicle or animal) collect numerous sensor readings while in transit. The nodes ultimately arrive back at a known rendezvous point (e.g., command center or animal pen), often as a large swarm and remain there for an extended period of time. The data stored on each sensor node is offloaded to a base station (BS) during this time. Moreover, since sensor nodes are typically powered by batteries with limited energy, energy harvesting techniques such as solar panel~\cite{sommer2013power} and Wireless Power Transfer (WPT)~\cite{xie2012making,krikidis2012rf} have investigated to extend lifetime of nodes. WPT is implementable by various technologies such as inductive coupling, magnetic resonate coupling, and electromagnetic radiation, for short, medium, and long distance applications, respectively~\cite{park2015innovative,sample2011analysis}. Presently, the long distance WPT system has been studied to power a large number of devices distributed in a wide area~\cite{xiao2014wireless,huang2014enabling,zhang2013mimo}.  

A number of considerations make the data collection in wildlife tracking non-trivial. 
First, the number of nodes can be quite large (several hundreds) and while the nodes normally arrive back in large groups, their exact arrival sequence is often unknown. 
Second, during days with cloudy skies and adverse solar charging weather conditions, the amount of solar energy harvested is reduced. In addition, charging efficiency of WPT becomes very low when the node is far from the WPT transmitter (large-scale channel fading)~\cite{ju2014throughput} or encounters antenna orientation bias (shown in Section~\ref{testbed}). It is thus critical to collect more data from those nodes before their harvested energy is exhausted. 
Third, the wireless link quality of data transmission between each node and the BS may vary with time. Having a node transmit during instances when the channel quality is poor is likely to result in packet reception errors, which in turn would require retransmissions and thus increased energy expenditure. 
Fourth, data should be downloaded from the nodes in a fair way. In particular, the amount of data collected from each node should be greater than a certain application-specific threshold. This is important to maximise the accuracy of data analysis, e.g., in the context of mobility modelling and population characteristics for wildlife monitoring~\cite{zhao2015optimal,jurdak2015autonomous}. 

Conventional scheduling such as the one employed in IEEE 802.15.4~\cite{liu2014design,ieee2003standard} are based on First Come First Served (FCFS), which we refer to as \textit{batch processing}. Batch processing has limited performance in real-world conditions with irregular radio channels and limited bandwidth. Any node with poor link quality occupies the channel due to retransmissions, while the nodes with higher link quality have to wait. In addition, batch processing does not support data collection fairness, potentially downloading a large amount of data from a small subset of nodes.

\begin{figure}[htb]
\centering
\includegraphics[width=0.5\textwidth]{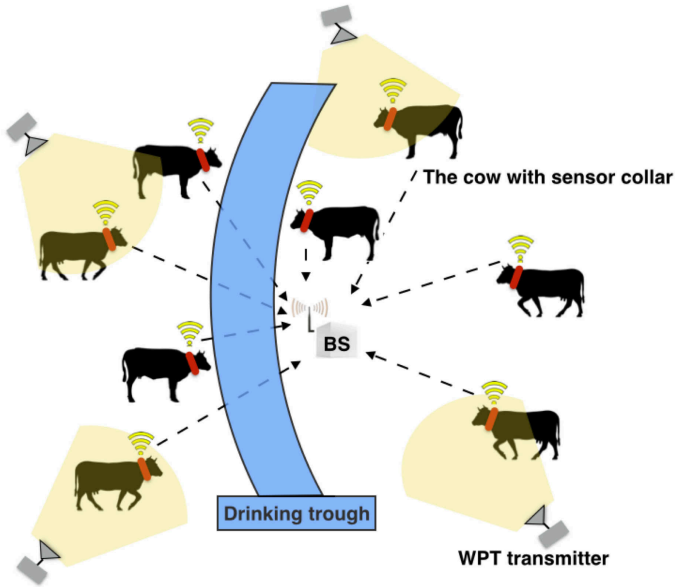}
\caption{Motivating Application: the energy harvesting mobile sensor network for cattle monitoring. The sensor collar is equipped with a solar panel and a WPT receiver.}
\label{fig_network}
\end{figure}

As an example, consider the problem of scheduling data transmissions in cattle monitoring application~\cite{Corke2010}. A sensor collar which contains embedded sensors (e.g., GPS, 3-axis accelerometer and magnetometer) is attached to the cow to record biological data~\cite{gonzalez2015behavioral,gonzalez2014wireless,dutta2014cattle}. The solar panel on the node harvests energy continuously during the sunny daytime. A WPT receiver on the sensor collar harvests energy from the WPT transmitter opportunistically when the animal stays in the charging range. The data is offloaded to a BS which is deployed near a cattle drinking trough. Figure~\ref{fig_network} depicts an energy harvesting mobile sensor network (MSN) for data collection. 

%Moreover, the application can be extended to other animals, i.e., flying foxes which typically swarm out in search of food at night and flock back to roosting camps during the daytime. A typical roosting camp can consist of hundreds of animals. The system aims to collect fine-grained spatiotemporal data about their movement patterns and environmental surroundings by attaching a sensor collar to these animals. 

In this paper, we propose Energy Harvesting Fair Scheduling (EHFS) to optimise data collection in a large-scale energy harvesting sensor network. The optimisation model schedules transmissions based on both link quality and residual energy of the node. It also ensures fairness by attempting to guarantee a certain application-specific amount of data collected from each node. We first show that this optimisation problem is NP-complete. Next, we propose EHFS heuristic algorithm to optimise the scheduling in linear time. The EHFS algorithm prioritises the nodes for scheduling based on a ratio of the link quality and harvested energy. This enables the nodes with the lowest energy reserves and good communication links to transfer their data first. In addition, we develop a state transition model to address the fairness criterion and maximise overall network goodput. 
Moreover, a Sensor-WPT testbed is built to characterise the WPT charging efficiency. Specifically, the experimental results show that WPT efficiency is jointly affected by distance between WPT transmitter and receiver, and their antenna orientation. 
While we use the wildlife monitoring application as a case study, the proposed optimisation model and EHFS algorithm are application-agnostic and hence applicable to a wide variety of large-scale energy harvesting mobile sensing scenarios with delay tolerance. 

The rest of paper is organised as follows: Section~\ref{relatedwork} presents related work on link scheduling and optimisation. We discuss the communication protocol on which EHFS is based in Section~\ref{protocol}. Section~\ref{model} formulates system and energy models in data collection. In Section~\ref{ehfsmodel}, we first present the scheduling optimisation and constraints. Then we prove that the optimisation problem is NP-complete and introduce our suboptimal algorithm. Section~\ref{evaluation} demonstrates the experiments on Sensor-WPT testbed, and compares the performance of the EHFS algorithm to the state-of-the-art in simulations. Finally, the paper is concluded in Section~\ref{conclusion}. 

\begin{figure*}[htb]
\centering
\includegraphics[width=1.0\textwidth]{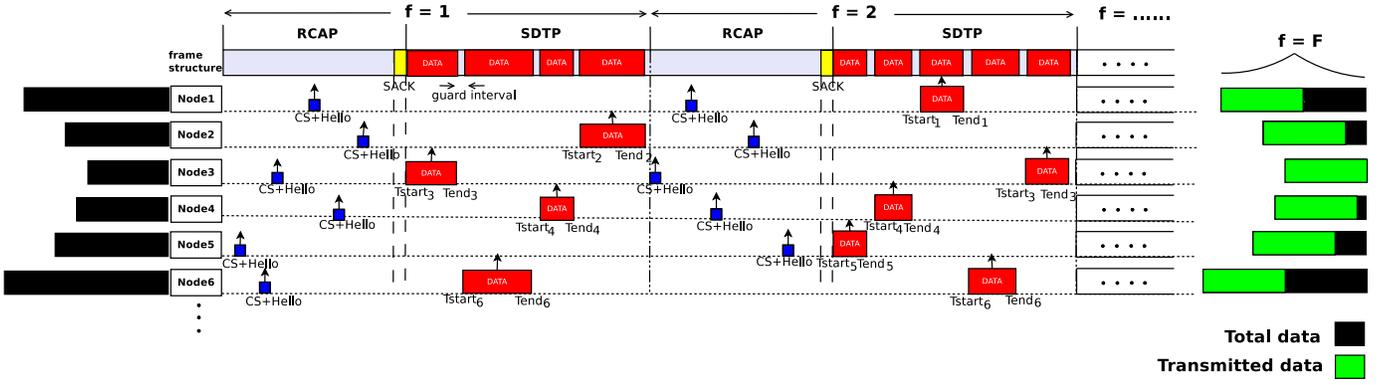}
\caption{The timing relationship. $Tstart_{i}$ and $Tend_{i}$ stand for the starting and ending time of node $i$'s data transmission respectively. The nodes with different data amounts are scheduled to transmit until $t = F$.}
\label{fig_channel}
\end{figure*}

\section{Related Work}
\label{relatedwork}
In this section, we review the literature on link scheduling and optimisation in wireless networks. To solve different optimisation goals, recent work considers throughput, energy consumption or time delay.

Extensive studies have been conducted on link scheduling in cellular networks. In \cite{schulman2010bartendr}, the link quality is predicted by an application framework which tracks the direction of travel of mobile phones at the BS. They develop energy-aware scheduling algorithms for different application workloads such as syncing or streaming. Some scheduling optimisations which consider multicast \cite{low2010optimized}, quality-of-service assurance \cite{wu2004downlink} and fair relaying with multiple antennas \cite{lin2012fair} are proposed to achieve optimal delay, capacity gain or network utility. 
The majority of related work has focused on addressing the scheduling problem in the context of wireless networks~\cite{fooladivanda2013joint,wu2013flashlinq,zhou2010distributed,ji2013delay}. However, the notion of fairness in wireless networks focuses on fair allocation, such as channels, tasks among different queues, or time slots among the links in each super frame, which is different from the fairness in data collection of MSN. 

A link scheduling for maximum throughput-utility in single-hop networks with the constraint of network delay is presented in \cite{Neely2010}. It establishes a delay-based policy for utility optimisation. The policy provides deterministic worst-case delay bounds with total throughput-utility guarantee. The author in \cite{Opportunisticscheduling2011} proposes an opportunistic scheduling algorithm that guarantees a bounded worst case delay in single-hop wireless networks. However, those scheduling algorithms are not applicable in MSNs, because they do not consider the constraints of energy and fairness of collection. In \cite{shaojie2012}, a sensing scheduling among sensor nodes is presented to maximise the overall Quality of Monitoring utility subject to the energy usage. The scheduling algorithm maximises the overall utility which is to evaluate quality of sensor readings based on the greedy algorithm. For body sensor networks, Sidharth, \textit{et al.} focus on polling-based communication protocols, and address the problem of optimising the polling schedule to achieve minimal energy consumption and latency \cite{nabar2010minimizing}. They formulate the problem as a geometric program and solve it by convex optimisation.

To the best of our knowledge, there is no research focusing on link scheduling optimisation for fair data collection in energy harvesting MSNs. The recent work in the literature is not applicable because they do not optimise the scheduling with the requirements of both energy consumption and data reception fairness. The key difference of our work over previous scheduling optimisation is that for a single-hop MSN which includes a large number of energy harvesting nodes, data collection is maximised in a fair way. We formulate the transmission scheduling optimisation model in Section~\ref{ehfsmodel}.

\section{Communication Protocol}
\label{protocol}
In this section, we present a communication protocol to improve the data collection performance under our specific constraints. 

We propose a communication protocol for scheduling optimisation in MSN. We utilise a 2-stage communication model, with random channel access period (RCAP) followed by scheduled data transmission period (SDTP) (see Figure~\ref{fig_channel})~\cite{li2014kappa}. The two periods interchange periodically until all the nodes finish data transmissions. %Sensor nodes do not keep track of the schedule while away from the BS, they only participate when in the range of the BS. 

The purpose of the RCAP is to collect information about sensor nodes, including their current link quality, the amount of available data, and their power resources. This data fits in a single \textit{Hello} packet and the nodes compete for the channel in a random-access fashion. Nodes check the radio channel for other data transmissions by using carrier sensing (CS) to avoid packet collisions and the reception of \textit{Hello} packets is acknowledged by the BS, so the nodes can turn off their radios until the end of the RCAP. However, if \textit{Hello} packets collision happen, the senders have to back off a random time to sense the channel again.

The BS calculates the transmission schedule at the end of RCAP by running the EHFS algorithm that we illustrate in Section~\ref{heuristic}. BS informs all sensor nodes the optimal schedule by broadcasting a \textit{SACK} packet at the end of the RCAP. 

The SDTP is driven by the schedule calculated by the EHFS algorithm. The nodes find their transmission slot (\textit{DATA} slot) within the super frame and only transmit during their scheduled time to prevent interference.  The length of the DATA slots is selected by the scheduler and will typically allow for multiple packet transmissions. We use guard intervals to prevent packet collisions due to time-synchronisation errors. With a large number of nodes, some of them may fail to communicate with the BS during RCAP. However, these nodes consume limited energy due to a long sleeping time during the SDTP.

\section{System Model}
\label{model}
On the basis of Section~\ref{protocol}, the BS aggregates the nodes and channel information in the RCAP in order to schedule the transmissions. In this section, we explain the basic notations and present an abstract generalisable model of the network, which is used for the optimisation model presented in the Section~\ref{optmodel}. We assume that there are $N$ nodes that directly communicate with the BS using single-hop communication. %The nodes typically arrive in large groups but their exact arrival sequence is unknown. 
The residual energy of a node $i$ at the beginning of RCAP is denoted by $E_{i}^{0}$. In order to prevent a node from completely depleting its battery, we assume that a node powers down if the residual energy goes below a certain threshold $E_{td}$. 
In this paper, a node in such a state is referred to as a {\it dead node}. This may happen if any node consumes more energy than it harvests. 
The wireless channel between each node and the BS is typically influenced by a variety of environmental factors and the transmission noise. The channel variability in turn influences the Packet Reception Rate (PRR) of the node. We estimate the PRR as a function of empirically collected RSSI traces from a real testbed as outlined in Section~\ref{simulation}. 

\subsection{Channel Model for Data Transmission}
According to the super frame as shown in Figure~\ref{fig_channel}, we divide the SDTP to a number of slots $S$, where, $S = \sum_{i=1}^{N} \Delta T_{i}$. Time slot $j$ ($j \in [1, S]$) is allocated by the BS to only one node's transmission for the purpose of avoiding collisions. Therefore, the allocated time $\Delta T_{i}$ of the node $i$ contains multiple time slots in one super frame. EHFS calculates optimal solutions for the nodes in each frame so that the schedule is optimised globally. $F$ is defined as the total number of super frames needed for all the nodes to finish their data transmissions. The sequence number of super frame is denoted as $f$ ($f \in [1,F]$). 
We assume the residual energy when node $i$ arrives at the data collection centre is $E_{i}^{0} (i \in [1, N])$. The PRR is indicated by $q_{i}^{f}$, where $q_{i}^{f} \in [0, 1]$. Additionally, $q_{i}^{f}$ may change from one super frame to the next due to the time-varying channel. We assume $q_{i}^{f}$ does not change during the super frame due to block fading. %since the flying foxes are not highly mobile in the data collection centre. 
The path loss of the sensor-BS channel can be approximated as free-space path loss \cite{goldsmith2005wireless} and is given by, 
\begin{equation}
L(d_{i,BS}) = K_{1}(d^{f}_{i,BS})^{K_{2}}, 
\label{eq_L}
\end{equation} 
where $K_{2}$ indicates the path loss component. $d_{i,BS}^{f}$ is the distance between the node $i$ and BS at frame $f$. $K_{1}$ is denoted by 
\begin{equation}
K_{1} = \frac{(4 \pi)^2}{G_{tx}G_{rx}\lambda_{0}^{2}}, 
\label{eq_K1}
\end{equation} 
where $G_{tx}$ and $G_{rx}$ are the antenna gains of the transmitter and receiver, respectively. $\lambda_{0} = c/f_{0}$, which is a ratio of speed of light $c$ and carrier frequency $f_{0}$. 
We define Signal-to-Noise ratio (SNR) for data communication between the node and BS as $\gamma^{\prime}_{i}$. Given an additive white Gaussian noise (AWGN) with power $N_{0}$, 
\begin{equation}
\gamma^{\prime}_{i} = \frac{|\hslash|^2 P^{tx}_{i}}{N_{0}L(d_{i,BS}^{f})}, 
\label{eq_gamma}
\end{equation} 
where $P^{tx}_{i}$ denotes the transmit power of the node $i$. The small-scale fading is indicated by $\hslash$. Then, the average SNR for the node $i$ is calculated by 
\begin{equation}
\overline{\gamma}^{\prime}_{i} = \frac{P^{tx}_{i}}{K_{1}N_{0}(d^{f}_{i,BS})^{K_{2}}}.
\label{eq_avg_snr}
\end{equation} 

In this paper, we derive the packet error probability of the channel between the sensor node and the BS based on its outage probability, which provides the lower bound of the packet error probability under an assumption of ideal coding and modulation. For illustration purpose, Rayleigh Block fading is considered \cite{tse2005fundamentals}. The channel coefficient remains constant within each block, and varies between blocks. At time $t$, the outage probability at the node $i$ is given by 
\begin{equation}
\text{Pr}(\gamma^{\prime}_{i} < \gamma_{0} ) = \int_0^{\gamma_{0}} p(\gamma^{\prime}_{i}) d(\gamma^{\prime}_{i}) = 1 - \text{exp}(\frac{\gamma_{0}}{\overline{\gamma}^{\prime}_{i}}), 
\label{eq_bep_first}
\end{equation} 
where $\gamma_{0}$ is the SNR threshold required for successful reception at the BS. 

Substituting Equation~(\ref{eq_avg_snr}) into Equation~(\ref{eq_bep_first}), the packet error probability at the BS can be given by 
\begin{equation}
\text{Pr}_{i,BS} = 1 - \text{exp}(-K_{src} \cdot (d^{f}_{i,BS})^{K_{2}}), 
\label{eq_pep_first}
\end{equation} 
\begin{equation}
K_{src} = \frac{K_{1} N_{0} \gamma_{0}}{P^{tx}_{i}}.
\label{eq_ksrc}
\end{equation}

Therefore, the $q_{i}^{f}$ can be 
\begin{equation}
q_{i}^{f} = \text{exp}(-K_{src} \cdot (d^{f}_{i,BS})^{K_{2}}). 
\label{eq_pep_first}
\end{equation} 

The data payload stored on each node is represented by $\lambda_{i}$ and the fairness coefficients is $\kappa$, where $\kappa \in (0,100\%]$. Thus, the data reception fairness ensures that the number of data packets the BS collects from each node is not less than $\kappa \cdot \lambda_{i}$. 
We define the boolean variable $x_{ij}^{f}$ as a transmission indicator for node $i \in [1, N]$ associated with the slot $j \in [1, S]$ in the super frame $f \in [1, F]$. $x_{ij}^{f} = 1$ means node $i$ has $j$th slot reserved for transmission in super frame $f$. 
The number of data packets received by the BS in a super frame is defined as $\gamma_{f}$, where
\begin{equation}
\gamma_{f} = \sum_{i=1}^{N} \sum_{j=1}^{S} x_{ij}^{f}\cdot q_{i}^{f}, (f \in [1,F])
\end{equation}
Similarly, for all super frames, the data received by the BS from any node $i$ is defined as $\alpha_{i}$, where
\begin{equation}
\alpha_{i} = \sum_{f=1}^{F} \sum_{j=1}^{S} x_{ij}^{f}\cdot q_{i}^{f}, (i \in [1,N])
\end{equation} 

\subsection{Energy Model}
\label{wptmodel}
The energy consumption of nodes arises from the transmissions in RCAP and SDTP as shown in Figure~\ref{fig_channel}. In this paper, we let $e_{tx-hello}$, $e_{rx-hack}$ and $e_{rx-sack}$ be the energy consumption of transmitting one \textit{Hello} packet, receiving one \textit{HACK} and one \textit{SACK} of the nodes, respectively. The $e_{tx}$ represents energy consumption of transmitting one data packet. Due to the tiny energy consumption of carrier sensing compared to transmitting and receiving packets \cite{ergen2004zigbee}, we neglect the same in our model. The energy consumption of node $i$ in the RCAP is $\check{E_{A}}$, where
\begin{equation}
\check{E_{A}} = e_{tx-hello} + e_{rx-hack} + e_{rx-sack}
\end{equation}
We next define $\check{E_{Di}}$ as the energy that node $i$ consumes on data transmission in all super frames, where
\begin{equation}
\check{E_{Di}} = \sum_{f=1}^{F}\sum_{j=1}^{S} x_{ij}^{f} \cdot e_{tx}, (i \in [1,N])
\end{equation}

For energy harvesting, the node may receive energy input from multiple sources, such as solar, vibration, thermal, or WPT. The total energy input for the node is the sum of energy harvested from these sources over time. In this paper, we focus on two energy harvesting sources, namely, solar and WPT, and elaborate further on them. 
The amount of harvested energy from WPT depends on the transmit power, wavelength of the RF signals and the distance between the RF energy source and the harvesting node. 
We define the transmit power of WPT as $P^{WPT}_{tx}$. The harvesting power of node $i$ at frame $f$ is $P_{i,f}$. Therefore, the power harvested from the WPT transmitter can be calculated as follows: 
\begin{equation}
P^{WPT}_{i,f} = \delta_{i}(d)\delta_{i}(\theta) P^{WPT}_{tx} |h_{i,f}|^{2}
\label{eq_wptpower}
\end{equation}
where $\delta_{i}(d) \in (0,1]$ is a constant indicating WPT efficiency factor given the distance between node $i$ and the charger. The other constant $\delta_{i}(\theta) \in (0,1]$ denotes WPT efficiency given the antenna alignment between node $i$ and the charger. $h_{i,f}$ is the WPT channel gain between node $i$ and the charger at frame $f$. 
Furthermore, we denote the power harvested from solar panel as $P^{solar}_{i,f}$. 

Given the time of WPT is $\tau_{i}$ and the solar charging duration is $\tau^{\prime}_{i}$, the harvested energy of sensor $i$ is given by 
\begin{equation}
\Delta E_{i,f} = \delta_{i}(d)\delta_{i}(\theta) (P^{WPT}_{tx} \tau_{i}) |h_{i,f}|^{2} + P^{solar}_{i,f} \tau^{\prime}_{i}
\label{eq_wptenergy}
\end{equation}

\section{Fair Scheduling with Energy Harvesting}
\label{ehfsmodel}
In this section, we first formulate fair scheduling optimisation under the constraints of fairness and energy harvesting. We show that the optimisation problem is NP-complete. Next, a heuristic algorithm, EHFS is proposed to approximate the optimal solution.

\subsection{Optimisation Formulation}
\label{optmodel}
\begin{small}
\begin{align}
& maximize \: \: \: \:\: \: \: \:\: \: \: \: \sum_{f=1}^{F} \gamma_{f} \nonumber \\
& subject \:\: to: \: \: \: E_{i}^{0} - \sum_{f=1}^{F}(\check{E_{A}} \cdot \varphi_{i}^{f} + \Delta E_{i,f}) - \check{E_{Di}} \geq E_{td}, \nonumber \\ 
& \: \: \: \: \: \: \: \: \: \: \:\: \: \: \: \: \: \: \: \: \: \: \: \: \: \: \: \: \: \: \:\: \: \: \: \: \: \: \: \: \:\: \: \: \: \: \: \: \: \: \:\: \: \: \: \: \: \: \: \: \: (i \in [1,N]) \label{cons1} \\
& \: \: \: \: \: \: \: \: \: \: \alpha_{i} \geq \kappa \cdot \lambda_{i}, \:\:\: (i \in [1,N], \: \kappa \in (0,1]) \label{cons2} \\
& \: \: \: \: \: \: \: \: \: \: \alpha_{i} \leq \lambda_{i}, \:\:\: (i \in [1,N]) \label{cons3} \\
%& \: \: \: \: \: \: \: \: \: \: |\alpha_{i} - \alpha_{i^{\prime}}| \leq \mu \cdot |(\lambda_{i} - \lambda_{i^{\prime}})|, \:\:\: (i, i^{\prime} \in [1,N], \: \mu \in (0,1]) \\
%& \: \: \: \: \: \: \: \: \: \: \:\: \: \: \: \: \: \: \: \: \: \: \: \: \: \: \: \: \: \: \:\: \: \: \: \: \: \: \: \: \:\: \: \: \: \: \: \: \: \: \:\: \: \: \: \: \: \: \: \: \: \lambda_{i} >= \lambda_{i^{\prime}}) \\
& \: \: \: \: \: \: \: \: \: \: x_{ij}^{f} \leq 1, \:\:\: (i \in [1,N], \: j\in [1,S], \: f\in [1,F]) \label{cons4} \\
& \: \: \: \: \: \: \: \: \: \: \sum_{i=1}^{N}x_{ij}^{f} \leq 1, \:\:\: (j\in [1,S], \: f\in [1,F]) \label{cons5} \\
& \: \: \: \: \: \: \: \: \: \: \lambda_{i} - \sum_{g=1}^{f}\sum_{w=1}^{j}x_{iw}^{g} \cdot q_{i}^{g} \geq v_{ij}^{f}, \:\:\: (i \in [1,N], \: j \in [1,S], \nonumber \\
& \: \: \: \: \: \: \: \: \: \: \:\: \: \: \: \: \: \: \: \: \: \: \: \: \: \: \: \: \: \: \:\: \: \: \: \: \: \: \: \: \:\: \: \: \: \: \: \: \: \: \:\: \: \: \: \: \: \: \: \: \:  f\in [1,F]) \label{cons6} \\ 
& \: \: \: \: \: \: \: \: \: \: v_{ij}^{f} \geq v_{ij^{\prime}}^{f}, \:\:\: (j^{\prime} \geq j, \: j\in [1,S]) \label{cons7} \\
& \: \: \: \: \: \: \: \: \: \: v_{ij}^{f} \geq v_{ij^{\prime}}^{g}, \:\:\: (g \geq f, j^{\prime} \geq j, j\in [1,S], \: f\in [1,F]) \label{cons8} \\
& \: \: \: \: \: \: \: \: \: \: \sum_{a=1}^{F-f} \varphi_{i}^{f+a} \leq v_{ij}^{f}, \:\:\: (i \in [1,N], \: j \in [1,S]) \label{cons9} \\
& \: \: \: \: \: \: \: \: \: \: x_{ij}^{f} \leq \varphi_{i}^{f}, \:\:\: (i \in [1,N], \: j\in [1,S], \: f\in [1,F]) \label{cons10} 
\end{align}
\end{small}
Based on the notations in the problem formulation, we formulate the EHFS for finding the optimal schedules as follows. Objective function of the optimisation model is to maximise $\gamma_{f}$ of all $F$ super frames. Constraint (\ref{cons1}) specifies the minimum remaining energy to be above $E_{td}$. 
A node stops accessing the channel after all its data has been transmitted or constraint (\ref{cons1}) is violated. Consequently, it does not waste energy in RCAP in subsequent super frames. For this purpose, $\varphi_{i}^{f}$ is defined as an indicator of RCAP in a super frame for the node. If the node $i$ does not compete for the channel in the RCAP of super frame $f$, $\varphi_{i}^{f}$ is equal to 0. $\sum_{f=1}^{F}(\check{E_{A}} \cdot \varphi_{i}^{f})$ indicates the energy consumption of the node in the RCAP of all super frames.
Constraint (\ref{cons2}) guarantees that the BS receives sufficient data packets to meet the fairness requirement. 
Constraint (\ref{cons3}) limits the value of $\alpha_{i}$ by the total payload $\lambda_{i}$. 
Constraints (\ref{cons4}) and (\ref{cons5}) specify that at any data transmission time slot only one node communicates with the BS to prevent transmission collisions. 

The only unknown is the total number of super frames during which a node is required to transmit. In other words, $\varphi_{i}^{f}$ is not known. To determine $\varphi_{i}^{f}$, we define a variable $v_{ij}^{f}$ for node $i$ at any slot $j$ of super frame $f$. 
Accordingly, constraint (\ref{cons6}) presents whether node $i$ has stopped the data transmission or not. $\sum_{g=1}^{f}\sum_{w=1}^{j}x_{iw}^{g} q_{i}^{g}$ is the total received packets until the current slot $j$ of super frame $f$. If the amount of data packets received from node $i$ matches the size of payloads $\lambda_{i}$, $v_{ij}^{f}$ is equal to 0. 
Constraints (\ref{cons7}) and (\ref{cons8}) ensure the future slots $j^{\prime}$ and super frames $g$ have $v_{ij}^{f} = 0$ if $\lambda_{i}$ packets have been received from node $i$. 
Constraint (\ref{cons9}) guarantees all $\varphi_{i}^{f}$ of the future super frames is 0 if $v_{ij}^{f} = 0$. As a result, the remaining energy of node $i$ which is restricted by the RCAP indicator $\varphi_{i}^{f}$ stops decreasing in constraint (\ref{cons1}).
Constraint (\ref{cons10}) ensures that the node $i$ stops data transmission if $\varphi_{i}^{f} = 0$.

\subsection{EHFS Algorithm}
\label{heuristic}
Maximising the collected data presented in Section~\ref{optmodel} is a typical 0-1 Multiple Knapsack Problem (MKP) \cite{martello1990knapsack}. We reduce an instance of a MKP to our scheduling optimisation problem by assigning $\Delta T_{i}$ to each knapsack. Therefore, the capacity of the knapsack is equal to $\Delta T_{i}$. The items to be put in knapsacks are data packets whose size is prorated by $q_{i}^{f}$. The parameters of the energy and fairness conditions (constraint (\ref{cons1}) and (\ref{cons2})) are chosen so that they are satisfied by any placement of items. In this way, optimal placement of items in knapsacks is reduced to such an instance of our scheduling problem. Since the problem is obviously an NP problem, this shows that our scheduling problem presented in the Section~\ref{optmodel} is NP-complete.

We propose a EHFS algorithm to approximate the optimal solution. Due to the prominent effect of energy harvesting and link quality variation on the scheduling, a ratio value for the node $i$ is denoted as $\eta_{i}^{f}$, where
\begin{equation}
\eta_{i}^{f} = \frac{q_{i}^{f}}{E_{i}^{f}}, \forall i \in [1,N], \forall f \in [1,F]
\end{equation}
Accordingly, $E_{i}^{f}$ is obtained by
\begin{equation}
E_{i}^{f} = E_{i}^{0} - \sum_{f^{\prime}=1}^{f}(\check{E_{A}} \cdot \varphi_{i}^{f^{\prime}} + \Delta E_{i,f^{\prime}}) - \sum_{f^{\prime}=1}^{f}\sum_{j=1}^{S} x_{ij}^{f^{\prime}} \cdot e_{tx}
\end{equation}
The motivation of calculating $\eta_{i}^{f}$ is to prioritise the nodes based on both the link quality and harvested energy. 
The EHFS algorithm gives a high transmission priority to the node with larger $\eta_{i}^{f}$. This method achieves large data reception because for the nodes with the same $q_{i}^{f}$, the node with the smallest $E_{i}^{f}$ gets higher transmitting priority. Similarly, for the nodes with the same $E_{i}^{f}$, one with higher $q_{i}^{f}$ has higher priority. 

In our algorithm, the node works in three states, Access \& Data transmission (AD), NonAccess (NA) and NonData (ND). In AD state, the node competes for the channel in RCAP and transmits data in SDTP as shown in Figure~\ref{fig_channel}. In NA state, the node neither accesses the channel nor transmits data but only receives the \textit{SACK} packets for the purpose of saving energy in the super frame. More importantly, none of the nodes, which are in the NA state transmit data given that no time slots are allocated to them. This helps more nodes achieve fairness. In ND state, the node does not turn on the radio and remains in sleep mode. Note that no matter which state the node works in it harvests energy by WPT transmitter. 

The EHFS algorithm develops two steps to maximise the data reception with $\eta_{i}^{f}$. It is implemented as shown in Algorithm~\ref{heu1}. 

\begin{algorithm}[t]
\begin{algorithmic}[1]
\begin{small}
\caption{EHFS Algorithm}
\label{heu1}
\STATE{nodes are in AD state and compete the channel}
\STATE{The BS calculates $\eta_{i}^{f}$ for the node $i$, $\forall f \in [1,F]$}
\STATE{The BS sorts the nodes by $\eta_{i}^{f}$, then $\eta_{i}^{f} \geq \eta_{i^{\prime}}^{f}, (i \neq i^{\prime}, i^{\prime} \in [1,N])$}
\STATE{The BS schedules the node $i$ to transmit}

\IF{$\alpha_{i} \geq (\kappa \cdot \lambda_{i})$}
\STATE{The node $i$ goes to NA state}
\STATE{The BS schedules the next one to transmit}
\ELSE
\STATE{The node $i$ remains in AD state}
\ENDIF

\IF{every node has $\alpha_{i} \geq (\kappa \cdot \lambda_{i}) \:\:\: \forall i \in [1,N]$}
\STATE{All the nodes transfer to AD state}
\STATE{The BS calculates $\eta_{i}^{f}$ for each node}
\STATE{The BS sorts the nodes by $\eta_{i}^{f}$, then $\eta_{i}^{f} \geq \eta_{i^{\prime}}^{f}, (i \neq i^{\prime}, i^{\prime} \in [1,N])$}
\IF{$E_{i} \geq E_{td}$}
\STATE{The BS schedules the node $i$ to transmit}
\ELSE
\STATE{The node $i$ changes state to the ND}
\STATE{The BS schedules the next one to transmit}
\ENDIF
\IF{$\alpha_{i} < \lambda_{i}$}
\STATE{The node $i$ remains in AD state}
\ELSE
\STATE{The node $i$ changes state to the ND}
\ENDIF
\ENDIF
\end{small}
\end{algorithmic}
\end{algorithm}

Initially, all nodes are in AD state and the BS schedules the node $i$ ($i \in [1,N]$) which has maximum $\eta_{i}^{f}$ to transmit data. The BS records the number of data packets from the node. 
Once the node $i$ meets the fairness of data reception (constraint (\ref{cons2})), it transfers to the NA state. The benefit of NA state is to reduce the channel competition since the number of nodes competing for the channel is decreased. Certainly, after the first step, all the nodes have at least $\kappa \cdot \lambda_{i}$ data packets being transmitted successfully and the fair reception of data is achieved. At the second step, all the nodes change the state from NA to AD. Then, the BS schedules the node with largest $\eta_{i}^{f}$ to transmit first. To maximise data reception, node $i$ remains in AD state until either constraint (\ref{cons1}) or (\ref{cons3}) no longer holds. Moreover, if the constraint of (\ref{cons1}) or (\ref{cons3}) is not fulfilled by the node $i$, it transitions to ND state. By using this approach, the number of data packets collected by the BS is maximised, meanwhile, the energy and fairness requirements are both achieved. 

\section{Performance Evaluation}
\label{evaluation}
Experiments are conducted on our Sensor-WPT testbed to measure the WPT efficiency as a fusion of distance between WPT transmitter and receiver and their antenna orientation. 
Then, given optimal schedules from the optimisation by AMPL, the performance of our EHFS algorithm is compared to the optimal schedules. We utilise empirical link quality, solar and WPT energy harvesting to evaluate the proposed algorithm in extensive simulations and compare its performance to state-of-the-art. 

\subsection{Experiments on Sensor-WPT Testbed}
\label{testbed}
We design two experiments to characterise the WPT efficiency factors $\delta_{i}(d)$ and $\delta_{i}(\theta)$, on our Sensor-WPT testbed. 

\begin{figure}[htb]
\centering
\includegraphics[width=0.5\textwidth]{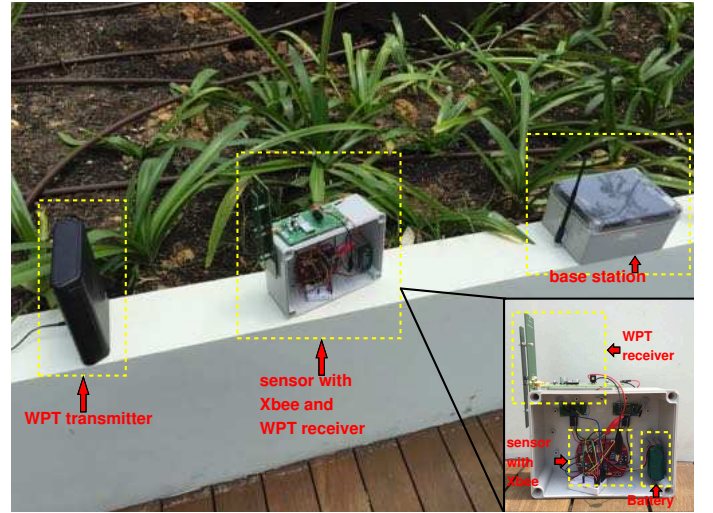}
\caption{The hardware setup contains the BS, WPT transmitter and sensor node with RFBee transceiver (for data transmission in 2.4 GHz) and WPT receiver.} 
\label{fig_hardware}
\end{figure}

In Sensor-WPT, each sensor node is equipped with a rechargeable battery and consumes energy on sensing and data transmission activities. A Powercast~\cite{wptTx} wireless charger transmits power to the sensor nodes by WPT. Since WPT charging is carried out in the 915 MHz band while sensor nodes communicate in the 2.4 GHz band, our network achieves simultaneous wireless information and power transfer. The isotropic radiated power of WPT transmitter with 8 dBi integrated antenna gain is 3W. The sensor node is connected to a P2110 powerharvesting board with a 1 dBi omni-directional antenna~\cite{P2110dataseet}. The hardware setup is shown in Figure~\ref{fig_hardware}. 

\begin{figure}[htb]
\centering
\includegraphics[height=1.5in,width=3.6in]{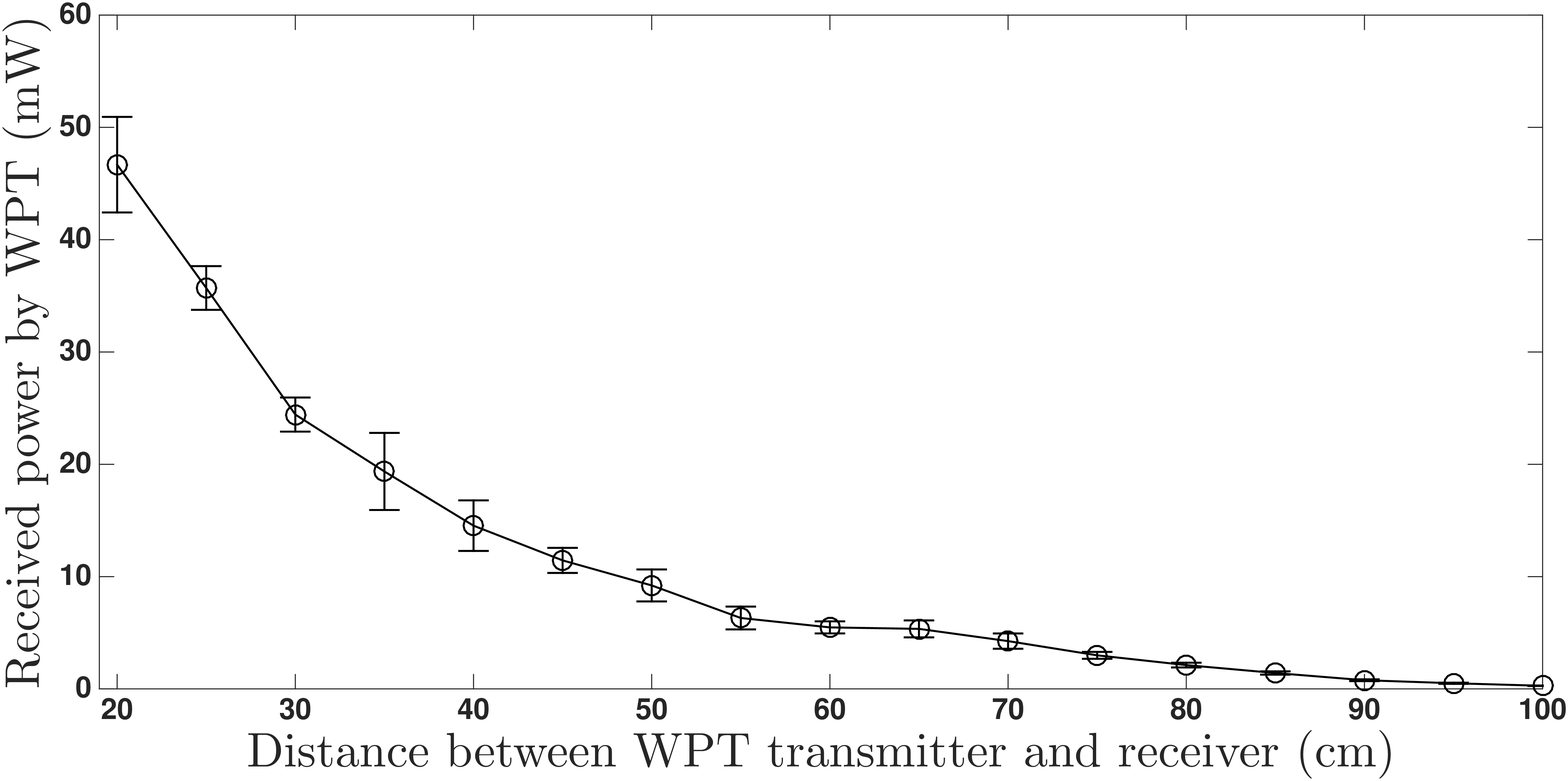}
\caption{Received power at the sensor by WPT. The error bars show the standard deviation over 250 packets.}
\label{fig_wpt_distance}
\end{figure}

In the first experiment, we measure the WPT efficiency factor on distance, $\delta_{i}(d)$ in Equation~(\ref{eq_wptenergy}). As shown in Figure~\ref{fig_wpt_distance}, the effective amount of power that can be captured by a sensor node varies with the distance between the node and the BS. Due to radiation exposure protection, the distance between WPT transmitter and P2110 powerharvesting board has to be further than 20 cm. 

\begin{figure}[htb]
\centering
\includegraphics[height=1.5in,width=3.6in]{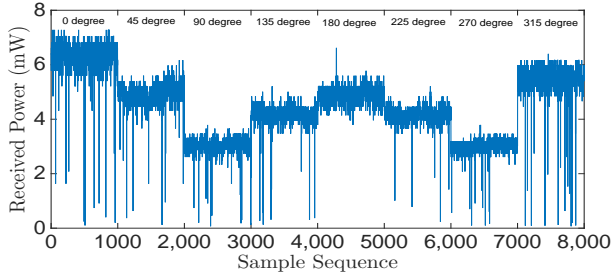}
\caption{Received power on the sensor node for varying antenna orientation of the WPT receiver.}
\label{fig_wpt_orientation}
\end{figure}

In the second experiment, we vary WPT receiver antenna orientation in order to configure $\delta_{i}(\theta)$ in Equation~(\ref{eq_wptenergy}). 
The distance between WPT transmitter and the sensor node is fixed at 55\textit{cm}. Initially, the antenna of WPT receiver on the node and WPT transmitter directly face towards each other. Therefore, the initial orientation is denoted as Zero degree rotation. The orientation increases 45 degrees every 1000\textit{s}, and the sensor node records 1000 samples at each orientation. The sensor node logs the sequence numbers and RSSI values of received packets in their flash. Figure~\ref{fig_wpt_orientation} shows that antenna orientation affects the received power at the WPT receiver. When the antenna orientation is at 90 and 270 degrees (two antennas are orthogonal to each other), the node harvests the lowest energy from the WPT transmitter.  


Based on the two experiments, we observe that the WPT efficiency is jointly affected by distance between WPT transmitter and receiver, and their antenna orientation. Finally, the parameters $\delta_{i}(\theta)$ and $\delta_{i}(d)$ are imported to our simulation configuration in the next Section. 

\subsection{Simulation Parameters}
\label{simulation}
The data collection network in the simulation contains one BS and $N$ nodes ($N \in [10, 300]$) which are randomly distributed within the open data collection centre. The node communicates with the BS using CC2420 radio in 2.4 GHz. The working temperature is measured as $25^{\circ}{\rm C}$, therefore, {\bf{$V_{cc}$}}, {\bf{$I_{tx}$}} and {\bf{$I_{rx}$}} is 3V, 35mA and 15mA, respectively~\cite{cc2420data}. We configure the remaining energy threshold of the sensor, {\bf{$E_{td}$}} to 1.67 mJ. 

\begin{figure}[htb]
\centering
\includegraphics[height=1.5in,width=3.6in]{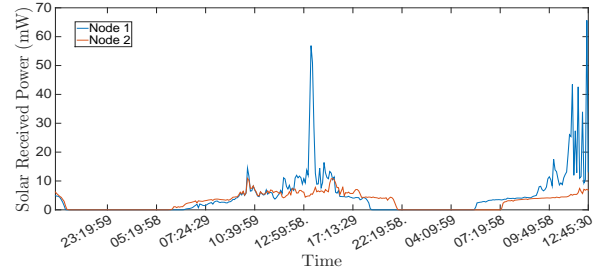}
\caption{The solar energy of two \textit{Camazotz} nodes over 43 hours.}
\label{elevel}
\end{figure}

Payload of the data packet has 32 bytes. The length of one \textit{Hello} packet is 10 bytes. Equally, \textit{HACK} and \textit{SACK} have the same length as \textit{Hello}. Therefore, we have
\begin{equation}
e_{tx-hello} = V_{cc} \cdot I_{tx} \cdot \frac{10 \times 8}{R_{b}} = 0.03 mJ
\end{equation}
\begin{equation}
e_{rx-hack} = e_{rx-sack} = V_{cc} \cdot I_{rx} \cdot \frac{10 \times 8}{R_{b}} = 0.01 mJ
\end{equation}
\begin{equation}
e_{tx} = V_{cc} \cdot I_{tx} \cdot \frac{32 \times 8}{R_{b}} = 0.1 mJ
\end{equation}
$E_{i}^{0}$ is given by a normal distribution with the mean value of 50 Joules according to the battery capacity of our sensors. The solar charging energy in the simulation makes use of the \textit{Camazotz} node, which has been developed for wildlife tracking~\cite{Jurdak2013}. \textit{Camazotz} reduces data sampling rate when the solar charge power is low. Figure~\ref{elevel} shows the harvested energy of the two nodes over 43 hours on wild flying foxes. It is observed that solar energy on the different nodes could be dynamic due to the mobility of nodes, weather, and landscape. Specifically, the empirical solar energy data is utilised in the energy model (shown in Section~\ref{wptmodel}) of our simulator. 

The WPT efficiency parameters, $\delta_{i}(\theta)$ and $\delta_{i}(d)$ are 0.5. Additionally, in our simulations, the value of $E_{i}^{0}$, $\delta_{i}(\theta)$ and $\delta_{i}(d)$ are given on purpose so that some dead nodes which run out of energy can be observed among different scheduling algorithms. %Figure~\ref{elevel} shows the values of $E_{i}^{0}$ when $N$ = 300 as an example. 
The RSSI trace recorded by the sensors in our testbed (shown in Figure~\ref{fig_hardware}) is imported to our simulator, which provides an environment to conduct repeatable simulations based on empirical data. %Figure~\ref{rssi} depicts a time segment which includes 1580 RSSI samples (The sampling rate of sensor is 3 samples per minute). 
In this paper, we convert the RSSI to PRR for the $q_{i}^{f}$ by the experimental results of PRR-RSSI relationship~\cite{Srinivasan2006}.

%\begin{figure}[htb]
%\centering
%\includegraphics[width=3.6in]{graph/rssi}
%\caption{RSSI trace of the sensor's data transmission.}
%\label{rssi}
%\end{figure}

\subsection{Scenarios and Metrics}
We simulate the EHFS algorithm in Node On Pasture (NOP) scenario and Node Arriving Pasture (NAP) scenario. In NOP scenario, we assume all the nodes are in the monitoring area from the start of experiment to the end. In NAP scenario, the nodes arrive at the area at different times. 
We evaluate three performance metrics: number of data packets received by the BS (data reception), the number of fair nodes and the number of dead nodes. Specifically, the \textit{fair node} denotes the node which fulfils $\alpha_{i} \geq \kappa \cdot \lambda_{i}$ (Constraint~(\ref{cons2})). We compare the performance of our EHFS algorithm with optimal solution at first. In NOP scenario, each node carries 80 KB data which is the payload generated by the sensor node. Since the number of nodes communicating with the BS in a short time is small in NAP scenario, we increase the data payload to 300 KB in order to explore the limits of the scheduling algorithms. For this reason, a node occupies the channel longer while more nodes enter the area in NAP scenario. 

To evaluate the performance of the EHFS algorithm in the NOP and NAP scenarios, two Greedy scheduling algorithms and FCFS algorithm are constructed in the numerical investigations. Because two basic elements used in the EHFS are the remaining energy represented by $E_{i}^{f}$ and link quality $q_{i}^{f}$ of node, the Greedy scheduling algorithms are formulated by them. The first Greedy algorithm is called Low Energy (LE) scheduling, namely, the transmission schedule is based solely on the $E_{i}^{f}$ of node. Lower $E_{i}^{f}$ implies higher priority of transmission at super frame $f$. High PRR (HP) scheduling is the second algorithm where the node with higher $q_{i}^{f}$ has higher priority. We compare them with the EHFS algorithm with $\kappa$ = 10\%, 50\% and 90\%. 

\subsection{Simulation Results}
\subsubsection{Comparing to Benchmark}
To compare to the optimal schedule shown in Section~\ref{optmodel}, we assess the performance of our algorithm when it operates in ten small-scale networks where the number of nodes is increased from 1 to 10. This initial comparison makes us aware of the performance difference between the optimal solution and our algorithm. The node $i$ carries 80 KB data, so $\lambda_{i} = 2500$. In fact, the comparison is not affected by different $\kappa$ values, thus we choose $\kappa$=50\% for both the optimal schedules and the EHFS algorithm. The optimal schedules achieve a maximum number of received data packets with the fairness and remaining energy constraints. They are constructed using AMPL and a state of the art ILP solver, Cplex 12.5, in a 2.7 GHz Intel core processor with 8 GB of memory. 

Table~\ref{ampl} summarises running time, the number of collected data packets and fair nodes. It is also found that there is no dead node in all tests and our algorithm guarantees exactly the same number of fair nodes as optimal schedules. On data reception, the EHFS algorithm and optimal solution have the maximum difference which is 706 when N = 9. On average, the number of packets in our algorithm is less than the AMPL output by around 1.16\%. Moreover, our algorithm is much more efficient than the optimisation model on runtime.

\begin{table}
    \centering
    \caption{Comparison between the optimal solutions and the EHFS algorithm}
    \begin{tabular} {|p{1.0cm}|p{1.3cm} p{1.3cm}|p{1.3cm} p{1.3cm}|} \hline
        \bf{Nodes} & \multicolumn{2}{c|}{\bf{AMPL (Cplex)}} &\multicolumn{2}{c|}{ \bf{EHFS }} \\ \hline
	\bf{} & \bf{Packets} & \bf{Runtime} & \bf{Packets} & \bf{Runtime} \\ \hline 
        	1 &  2499 & 1 s & 2491 & 0.07 s \\ \hline
	2 &  4999 & 12 s & 4981 & 0.1 s \\ \hline
        	3 &  7499 & 28 s & 7475 & 0.04 s \\ \hline
        	4 &  9998 & 63 s & 9954 & 0.06 s \\ \hline
        	5 &  12498 & 1 m 27 s & 12484 & 0.06 s \\ \hline 
        6 &  14998 & 5 m 15 s & 14465 & 0.06 s \\ \hline
        7 &  17498 & 1h 3 m & 17353 & 0.08 s \\ \hline
        8 &  19997 & 6 h 53 m & 19808 & 0.08 s \\ \hline
        9 &  22499 & 19 h 12 m & 21793 & 0.22 s \\ \hline
        10 &  24998 & 36 h 29 m & 24583 & 0.22 s \\ \hline
   \end{tabular}
\label{ampl}
\end{table}

\subsubsection{Node On Pasture Scenario}
\begin{figure}[htb]
\centering
\includegraphics[width=3.6in]{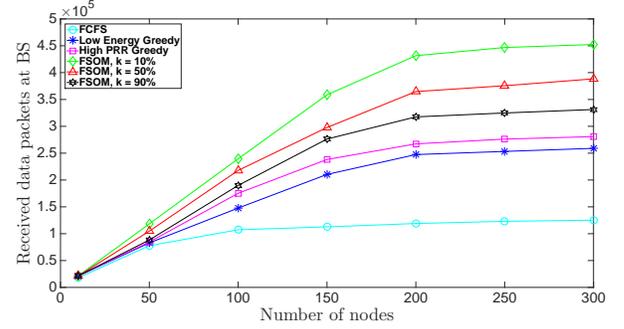}
\caption{Data packets collected by the BS, $N$ is from 10 to 300.}
\label{dynamic_datagraph}
\end{figure}

\begin{figure}[htb]
\centering
\includegraphics[width=3.6in]{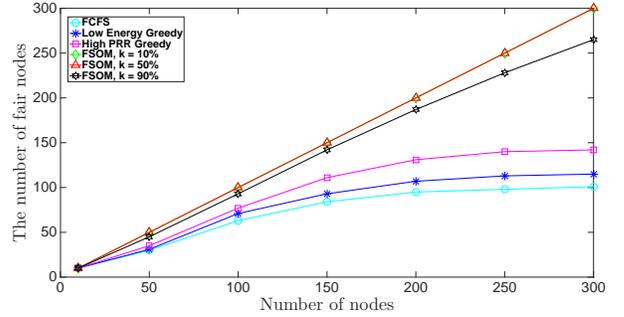}
\caption{Number of fair nodes among $N$, $N$ is from 10 to 300.}
\label{dynamic_fairnodes}
\end{figure}

Figure~\ref{dynamic_datagraph} and~\ref{dynamic_fairnodes} show the performance of the aforementioned four scheduling algorithms on the data reception and fairness. When there are only 10 nodes in the network, they have pretty similar performance. However, FCFS, LE and HP collect 75.6\%, 45.7\% and 41.3\% less data packets than our algorithm when N = 300. With WPT energy harvesting, it is observed that more data packets are collected with more nodes. The number of fair nodes of our algorithm is more than the ones of FCFS, LE and HP for 200, 180, 155 nodes when $\kappa$ = 50\% and N = 300. The reason is that LE scheduling fails when the low energy nodes have poor link quality. The nodes with high PRR are not scheduled, however, they still consume energy on channel competitions in RCAP. For HP scheduling, the nodes with high PRR occupy the SDTP for multiple super frames until they finish the transmissions. This leads to a large number of dead nodes. However, those nodes could have potentially gained higher data reception. In contrast, our algorithm makes the schedule based on $\eta_{i}^{f}$ which considers both remaining energy and link quality. Moreover, it also achieves the fairness of data collection. 

\begin{figure}[htb]
\centering
\includegraphics[width=3.6in]{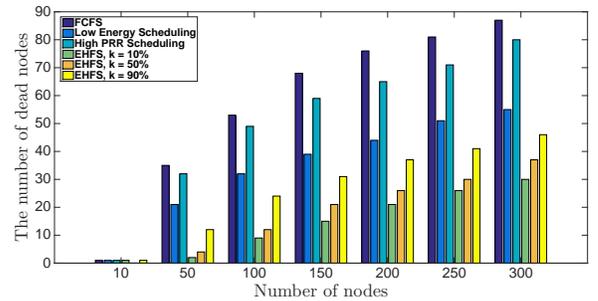}
\caption{Number of dead nodes among $N$, $N$ is from 10 to 300.}
\label{dynamic_deadnodes}
\end{figure}

We find the data reception and fair nodes of FCFS, LE and HP do not vary significantly from N = 150 to 300. The reason is indicated by dead nodes which are shown in Figure~\ref{dynamic_deadnodes}. It shows FCFS, LE and HP have much more dead nodes than the EHFS algorithm starting from N = 50. 

According to the EHFS algorithm, we know that $\kappa$ is a crucial variable which affects the states transition of node $i$. The performance of our algorithm varies with different $\kappa$ value. As shown, they are similar for $\kappa$ = 10\%, 50\% and 90\% when N is 10. From N = 50 to N = 300, $\kappa$ = 10\% performs better than 50\% and 90\%. The reason is that any node which is scheduled to transmit occupies more super frames when $\kappa$ is increased due to the fairness constraint (\ref{cons2}). It makes the other nodes compete the channel in RCAP repeatedly and cost energy. However, increasing $\kappa$ achieves more data collected from the single node, which benefits some application for individual sensor monitoring. Therefore, the configuration of $\kappa$ depends on the application requirement. 

\subsubsection{Fairness Parameter Effect}
\begin{figure}[htb]
\centering
\includegraphics[width=3.5in]{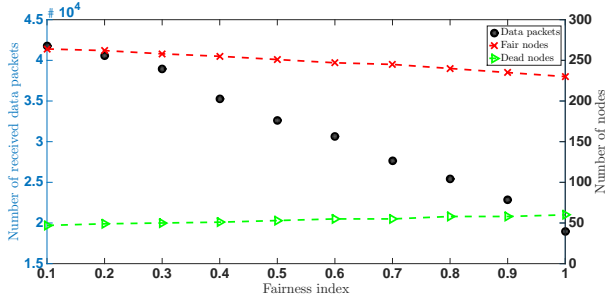}
\caption{The effect of $\kappa$ on the performance of EHFS algorithm. $N$ = 300 and $\kappa$ is from 10\% to 100\%.}
\label{optfair}
\end{figure}

Based on the preceding simulations, it is observed that different $\kappa$ affects the performance of our algorithm. Essentially, the $\kappa$ decides the fairness level in EHFS. In this experiment, we analyse the impact of $\kappa$ in the NOP scenario with 300 nodes. Specifically, the $\kappa$ is varied from 10\% to 100\%. The performance of data reception, fair nodes and dead nodes are shown in Figure~\ref{optfair}.

As shown in Figure~\ref{optfair}, data reception rate decreases and the number of dead nodes slowly increases with the increasing $\kappa$. This is because the transmission duration of one node is extended when $\kappa$ is increased. Other nodes with small harvested energy $\Delta E_{i,f}$ deplete their energy due to RCAP if the channel is occupied by someone with high $\eta_{i}^{f}$ for a long time. Their data is not collected by the BS before the nodes exhaust the energy. As observed, energy harvesting can only retard this energy depletion instead of addressing it thoroughly since the charging efficiency is affected by the environmental factors. We also find that the scheduling with smaller $\kappa$ achieves larger number of fair nodes. 

Therefore, Figure~\ref{optfair} indicates a tradeoff, namely, higher $\kappa$ guarantees more data packets collected from individual node while sacrificing the system throughput; smaller $\kappa$ achieves a higher system throughput, however, it does not guarantee most of data can be collected from individual node since the BS gives the priority to the one with larger $\eta_{i}^{f}$ after all nodes satisfy the fairness constraint. 
%$\kappa$ changed from 40\% to 50\% keeps a balance between the data reception from each node and total number of dead nodes.

\subsubsection{Node Arriving Pasture Scenario}
In this set of experiments, we test the scheduling algorithms when nodes arrive at the data collecting point with a specific arrival rate. We assume the inter-arrival time of nodes is exponentially distributed which is typically used to model situations involving the random time between arrivals to a service facility \cite{willkomm2009}. 

\begin{figure}[htb]
\centering
\includegraphics[width=3.6in]{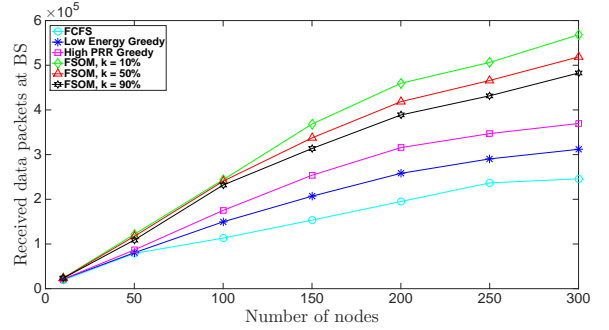}
\caption{Data packets collected by the BS. $N$ is from 10 to 300.}
\label{lambda_datagraph}
\end{figure}

\begin{figure}[htb]
\centering
\includegraphics[width=3.6in]{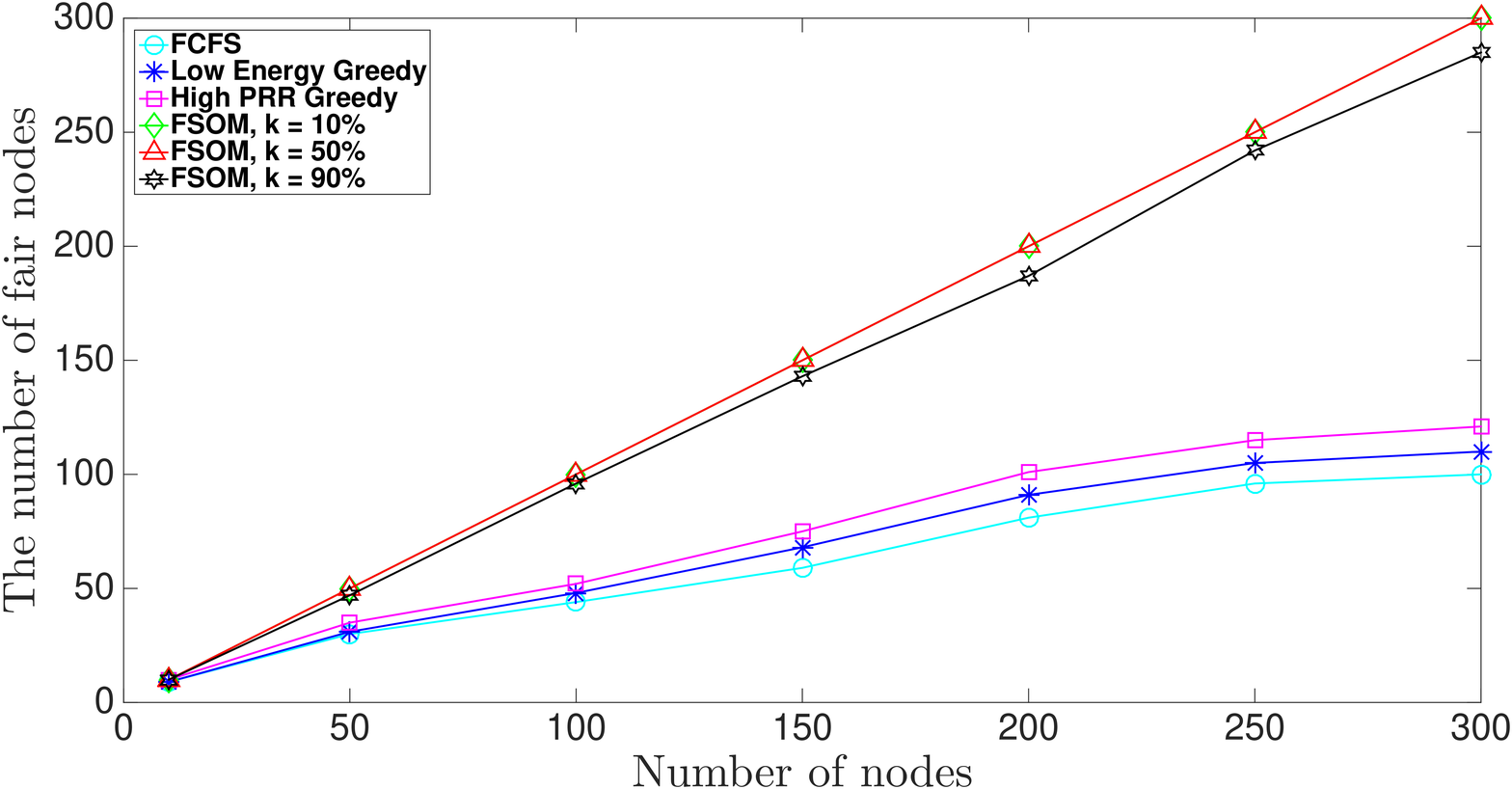}
\caption{Number of fair nodes among $N$, $N$ is from 10 to 300.}
\label{lambda_fairnodes}
\end{figure}

\begin{figure}[htb]
\centering
\includegraphics[width=3.6in]{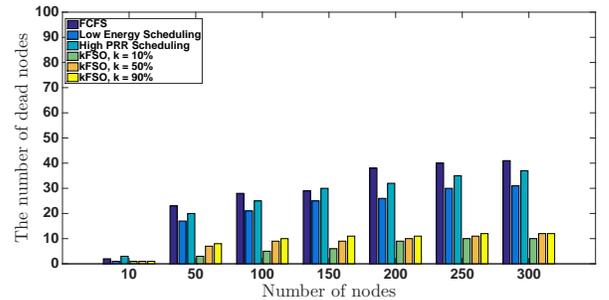}
\caption{Number of dead nodes among $N$, $N$ is from 10 to 300.}
\label{lambda_deadnodes}
\end{figure}

From Figure~\ref{lambda_datagraph} we find that the EHFS algorithm has up to 2.3 times as many collected data packets as FCFS. It outperforms LE and HP by nearly 1.7 times as well. The reason is the newly arrived nodes fail to transmit since the transmitting node have not finished the transmission due to retransmissions. From Figure~\ref{lambda_fairnodes}, we observe the difference of fairness which is achieved by different $\kappa$ is smaller than the one in NOP scenario. That is because the BS schedules a small number of nodes in one super frame in NAP scenario. The first step of EHFS algorithm is completed faster, hence more nodes achieve fairness in NAP scenario. Likewise, the number of dead nodes in our algorithm has small difference in Figure~\ref{lambda_deadnodes}. Due to the increase of $\lambda_{i}$ in this application, there are 12 dead nodes with the $\kappa$ = 90\% in our algorithm at the maximum. 
Moreover, in Figure~\ref{lambda_deadnodes}, the FCFS, LE and HP also have smaller dead nodes compared with the NOP scenario. The reason is that a small number of nodes is scheduled to transmit at one super frame and they can finish 300 KB data transmission soon. So the newly arrived nodes have small channel competition. 

\subsubsection{WPT Efficiency Effect}
\begin{figure}[htb]
\centering
\includegraphics[width=3.6in]{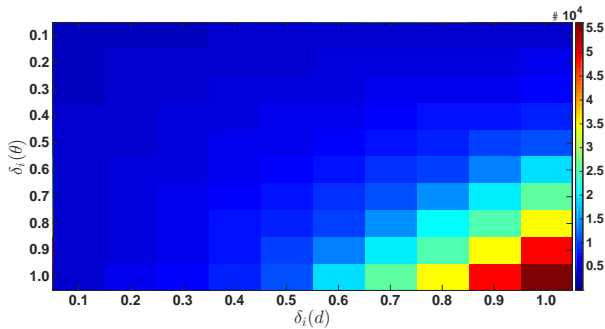}
\caption{Data packets reception according to WPT antenna orientation $\delta_{i}(\theta)$ and distance $\delta_{i}(d)$. $N = 50$ and $\kappa = 50\%$.}
\label{wpt_disorientation}
\end{figure}

According to Equation~(\ref{eq_wptenergy}) and the experiments in Section~\ref{testbed}, it is observed that $\delta_{i}(\theta)$ and $\delta_{i}(d)$ jointly affect harvested energy $\Delta E_{i,f}$ of the sensor node and the performance of scheduling algorithm. Figure~\ref{wpt_disorientation} illustrates the impact of the WPT efficiency on the data packets reception of EHFS given that the number of nodes is 50 and $\kappa$ is 50\%. Data reception increases by increasing $\delta_{i}(\theta)$ and $\delta_{i}(d)$ since the nodes harvest more energy via WPT. Specifically, when the nodes are close to the WPT transmitter ($\delta_{i}(d) = 1$) with WPT receiver antenna alignment ($\delta_{i}(\theta) = 1$), the data reception has the maximum value which is about 56250 packets. Even in the worst case ($\delta_{i}(d), \delta_{i}(\theta) = 0.1$), EHFS algorithm can still achieve the reception of 3251 packets. 

\section{Conclusion}
\label{conclusion}
In this paper, we have proposed and evaluated a fair link scheduling optimisation model with the objective of maximising the data reception in the data collection of energy harvesting MSN. The super frame structure is developed for the BS to collect data from the sensor nodes. We have proved that the scheduling optimisation is an NP-complete problem. Therefore, the EHFS algorithm is proposed to approximate the optimal solutions in polynomial time. Our algorithm schedules the transmissions of the nodes based on $\eta_{i}^{f}$ and three working states in two steps. 
With the wildlife monitoring application and our Sensor-WPT testbed, we have shown the numerical performance of the EHFS algorithm based on the solar energy, WPT charging efficiency, and RSSI. We have compared our algorithm with the optimal schedules of the optimisation model and presented extensive simulations incorporating both node on the pasture and node arriving pasture scenario. Specifically, the EHFS algorithm provides a near-optimal scheduling to the data collection in the energy harvesting MSN. 

\ifCLASSOPTIONcaptionsoff
  \newpage
\fi

\bibliographystyle{IEEEtran}
\bibliography{EHFS}

\end{document}